# Electronic states and magnetic response of $MnBi_2Te_4$ by scanning tunneling microscopy and spectroscopy

Yonghao Yuan[1,2,†], Xintong Wang[1,†], Hao Li[3,4,†], Jiaheng Li[1,†], Yu Ji[1], Zhenqi Hao[1], Yang Wu[5], Ke He[1,2], Yayu Wang[1,2], Yong Xu[1,2], Wenhui Duan[1,2], Wei Li[1,2,*] and Qi-Kun Xue[1,2,*]

[1]*State Key Laboratory of Low-Dimensional Quantum Physics, Department of Physics, Tsinghua University, Beijing 100084, P. R. China*

[2]*Frontier Science Center for Quantum Information, Beijing 100084, China*

[3]*School of Materials Science and Engineering, Tsinghua University, Beijing, 100084, P. R. China*

[4]*Tsinghua-Foxconn Nanotechnology Research Center, Department of Physics, Tsinghua University, Beijing 100084, P. R. China*

[5]*Department of Mechanical Engineering, Tsinghua University, Beijing, 100084, P. R. China*

†These authors contributed equally to this work.

*Wei Li Email: weili83@tsinghua.edu.cn Phone: +8610-62795838

*Qi-Kun Xue Email: qkxue@mail.tsinghua.edu.cn Phone: +8610-62795618

**ABSTRACT**

Exotic quantum phenomena have been demonstrated in recently discovered intrinsic magnetic topological insulator $MnBi_2Te_4$. At its two-dimensional limit, quantum anomalous Hall (QAH) effect and axion insulator state are observed in odd and even layers of $MnBi_2Te_4$, respectively. The measured band structures exhibit intriguing and complex properties. Here we employ low-temperature scanning tunneling microscopy to study its surface states and magnetic response. The quasiparticle interference patterns indicate that the electronic structures on the topmost layer of $MnBi_2Te_4$ is different from that of the expected out-of-plane A-type antiferromagnetic phase. The topological surface states may be embedded in deeper layers beneath the topmost surface. Such novel electronic structure presumably related to the modification of crystalline structure during sample cleaving and re-orientation of magnetic moment of Mn atoms near the surface. Mn dopants substituted at the Bi site on the second atomic layer are observed. The ratio of Mn/Bi substitutions is 5%. The electronic structures are fluctuating at atomic scale on the surface, which can affect the magnetism of $MnBi_2Te_4$. Our findings shed new lights on the magnetic property of $MnBi_2Te_4$ and thus the design of magnetic topological insulators.

**INTRODUCTION**

Introducing magnetism to a time-reversal-invariant topological insulator may generate exotic quantum phenomena(*1-15*), such as the quantum anomalous Hall (QAH) effect and axion insulator state. The QAH effect has been realized in dilute magnetic topological insulator Cr and V-doped $(Bi, Sb)_2Te_3$(*5-10*). The magnetic doping, which affects the local magnetism and chemical potential(*16*), might give rise to the low observing temperature of the quantization. An intrinsic magnetic topological insulator without magnetic dopants is ideal to realize QAH effect at higher temperature. $MnBi_2Te_4$, a layered stoichiometric material, which can be viewed as intergrowth of rock-salt structure MnTe within the quintuple layer of topological insulator $Bi_2Te_3$ (Fig. 1A), is a material of such kind(*17-19*). Quantized Hall resistances have been recently observed in $MnBi_2Te_4$ at two-dimensional (2D) limit(*20-22*). However, several properties of this newly discovered material are still unclear, which are important to further improve the performance of the material. Its band structure measured by angle-resolved photon emission spectroscopy (ARPES) shows deviations from the calculated results(*17-19, 23-28*). For example, the observations of magnetism-induced gap opening at the Dirac point are either too large or too small (absent)(*17-19, 23-28*) and the assignment of the topological surface states are still controversial(*26-28*). Here we employ scanning tunneling microscopy (STM) to systematically study the electronic structure of MnBi$_2$Te$_4$ in both real and momentum spaces, and the complementary information helps us to have a better understanding of the material.

**RESULTS**

Fig. 1B shows a typical STM topographic image of MnBi$_2$Te$_4$. The observed hexagonal lattice structure here corresponds to the Te-terminated (111) surface. The schematic lattice structure of MnBi$_2$Te$_4$ is shown in Fig. 1A. Considerable defects are revealed in Fig. 1B. The defects can be grouped into two primary types, bright dots and dark triangles. The bright dots could be adatoms formed during sample cleavage process. The dominant defects are the dark triangles, similar to that in Cr doped (Bi, Sb)$_2$Te$_3$(*29*). They probably are Mn-on-Bi antisite defects (Mn$_{Bi}$) in the second atomic layer, denoted by the red arrow in Fig. 1A. The ratio of Mn/Bi substitutions is 5%. The scanning tunneling spectroscopy (STS) detects the differential tunneling conductance d$I$/d$V$ (inset of Fig. 1B), which provides a measure of the local density of states (DOS) of quasiparticles. The DOS over the energy range of -0.5 eV to the Fermi level ($E_F$) are quite low. Since obvious DOS minimum of the STS is hardly to be defined, location of the original Dirac point as well as the expected magnetism-induced gap opening at the Dirac point is undetermined in our results.

To investigate the electronic structures, we perform d$I$/d$V$ maps from the same area at various bias voltages ranging from -100 meV to 400 meV (Fig. 2A1-F1). Fig. 2A2-F2 show the corresponding fast Fourier transformation (FFT) results of the d$I$/d$V$ maps, in which the scattering wave vectors $\vec{q}_{QPI}$ in momentum space are extracted. The FFT shows spot features with six-fold symmetry along the $\overline{\Gamma}$-$\overline{M}$ direction of the surface Brillouin zone. In contrast to the bright and sharp FFT spots obtained from the surface states of Bi$_2$Te$_3$(*30*), the spot features here are quite broad and faint. In Bi$_2$Te$_3$, the QPI are mainly contributed by surface states(*30*), which are non-dispersive along the $k_z$-direction, leading to much clearer and sharper FFT spots.

Since the time-reversal symmetry is broken in out-of-plane A-type antiferromagnetic MnBi$_2$Te$_4$, the calculated constant-energy contour (CEC) of the surface states shows three-fold

symmetry(*19*) (see the blue shape in Fig. 2G). Then the dominant scattering wave vector $\vec{q}_s$ should be the one connecting the parallel segments of the CEC (see the double-headed red arrows). However, such a $\vec{q}_s$ expected from the scattering of surface states is oriented along the $\overline{\Gamma}$-$\overline{K}$ direction, which is inconsistent with the observed QPI results (where the $\vec{q}_{QPI}$ is along the $\overline{\Gamma}$-$\overline{M}$ direction).

A possible explanation for the above observations is as follows: 1. The bulk band scattering process mainly contributes to QPI patterns, giving rise to the broad and faint FFT spots. 2. The wave function of surface states is actually embedded in the deeper atomic layers of the sample and cannot be detected by surface-sensitive STM. The origin for the embedded surface states might be related to the modification of crystalline structure on the topmost layer during the sample cleaving process. The detected signal of STM exponentially decays with the increasing distance between the sample and the tip, therefore, the embedded surface states, which is fewer angstroms beneath the topmost layer, is missed by STM. No terrace-edge induced standing wave is observed by STM (Fig. 2I1, 2I2), which further support our speculation of the embedded surface states. Such a picture is also compatible with ARPES results. The Dirac surface bands are only visible in ARPES data taken with low photon energy (7 eV), by which the electronic structures of deeper layers are detected(*26, 27*). Note that the samples measured in ref. 27 and in our STM study are from the same batch. Another possibility to obtain faint FFT spots is that the electron scattering process occurs in a strongly correlated system, which is not the case in $MnBi_2Te_4$.

Therefore, to understand the origin of the FFT results, we further include the contribution of the bulk states together with surface states of the out-of-plane A-type antiferromagnetic $MnBi_2Te_4$. The corresponding calculated CECs (including both bulk and surface bands) of out-of-plane A-type antiferromagnetic phase (or AFM-*z*) are shown in the insets of Fig. 3B-D. However, the simulation results of self-auto-correlation of these CECs (Fig. 3B-D) patterns are inconsistent with experimental results (Fig. 2A2-F2). In order to make a clearer comparison, the scattering wave vectors obtained from the experimental results in Fig. 2H are added to the simulated results here as red arrows. The features at $\vec{q}_{QPI}$ are obviously not the dominant ones in Fig. 3B-D. Since STM detects the topmost layer of the material, we conclude that the electronic structures on the topmost layer of $MnBi_2Te_4$ are different from that of the expected out-of-plane A-type antiferromagnetic phase MnBi$_2$Te$_4$.

The deviation of the electronic structures here might originate from the re-orientation of magnetic moment of Mn atoms on the topmost layer. The translational symmetry along the $z$ direction is broken on the cleaved surface. It is natural to imagine that both the magnetic coupling strength and the orientation of the magnetic moments of Mn atoms are changed as they are closer to the surface. In addition, the cleaving process may also modify the magnetism as well as the crystalline structure of the surface. Such surface magnetism variation has already been found in CrI$_3$, whose bulk magnetism resembles to that of MnBi$_2$Te$_4$(*31*). The widely distributed Mn$_{Bi}$ dopants (defects) may also change the magnetism of MnBi$_2$Te$_4$. The variation of the surface magnetism is also proposed based on ARPES observations of gapless Dirac cone structure on MnBi$_2$Te$_4$(*26-28*), in which in-plane or disordered magnetic moments might appear on the surface.

We next try to check the possible magnetic structure of the top layers of MnBi$_2$Te$_4$. Fig. 3E, I show the schematics of another two possible magnetic structures, the in-plane A-type antiferromagnetism (AFM-*x*) and out-of-plane ferromagnetism (FM-*z*), respectively, in which the bulk and surface band structures are significantly different from that of the expected out-of-plane A-type antiferromagnetic MnBi$_2$Te$_4$, therefore might give rise to the observed FFT results. However, as shown in Fig. 3F-H and Fig. 3J-L, neither of their corresponding self-auto-correlation are consistent with the observed FFT results. Therefore, none of the above three magnetic structures is coincident with the QPI results. The magnetism of the surface of MnBi$_2$Te$_4$ still needs further investigation.

Fig. 2H summarizes the response of QPI wave vectors to magnetic fields and temperatures. Previous studies demonstrate that the antiferromagnetic transition temperature of MnBi$_2$Te$_4$ is ~ 25 K(*23, 24, 32-35*), and the magnetic order is aligned under magnetic field at 7.7 T(*23, 24*). Since the STM detects the top layers with re-oriented magnetic moments and its bulk states are dominant in the tunneling spectra, the QPI patterns are not changed neither under the magnetic field up to 12 T nor at the elevated temperature to 30 K (Fig. 2H). The observed Dirac point from ARPES(*26, 27*) are located at $E_D$ = -270 meV below $E_F$. However, around $E_D$, the STM QPI patterns are rather weak and the scattering wave vectors cannot be extracted from their FFT results. Therefore, no changes have been observed at 9 T in the QPI patterns around $E_D$.

We further examine the response of d$I$/d$V$ spectra to magnetic field. Magnetic fields show almost no observable effects on the large-bias-range d$I$/d$V$ spectra in Fig. 4A, which presents contrast to the emergence of Landau levels established on the surface states of topological insulators(*36, 37*), but is consistent with our picture that the topmost layer actually only contribute bulk band states. On the other hand, the small-bias-range spectra show that DOS near $E_F$ are fluctuating under magnetic fields (Fig. 4B, C). But no systematic evolution, such as gap opening or Zeeman splitting, is observed. The tunneling spectra show a dip feature at $E_F$, and its origin is still unclear.

Based on our STM observation, we would like to discuss the surface structure of MnBi$_2$Te$_4$. Generally, after sample cleavage, the new surface of bulk MnBi$_2$Te$_4$ is divided into two parts (layers) according to the depth from the topmost layer (see schematic in Fig. 4D): (*i*) the top layer with the thickness of few nanometers, which hosts an embedded Dirac cone structure due to the unclear magnetic order(*26, 27, 38*); (*ii*) the deep layer with the predicted out-of-plane antiferromagnetic order(*18, 19*). Depending on the detection depths of different techniques, the top layer is detectable to surface sensitive techniques, such as STM and ARPES, while the deep intrinsic layer is detectable to bulk sensitive techniques such as transport measurements.

Besides the layer separation of the sample, the existence of Mn$_{Bi}$ defects is another key factor to affect the ARPES and transport results. Since they are randomly distributed in the sample and their crucial role to the magnetism, we turn to discuss the dopant effect on electronic structures on two different spatial scales. Fig. 5A shows three kinds of averaged tunneling spectra taken on two types of defects and defect-free regions respectively, over an area of 25 nm × 25 nm. The peak positions of those averaged spectra (denoted by black arrows) are almost identical, although the intensities of the peaks are slightly different. Limited by the spot size of light, ARPES experiment obtains the spatial averaged signals, therefore the identical peak position here is consistent with the sharp band structure observed by ARPES. However, at atomic scale, Fig. 5B shows the spectra taken along the white dotted line in the inset of Fig. 5A, in which the electronic structure exhibits large fluctuating near $E_F$ (see the red arrows and the red parts of the spectra). The fluctuating DOS might affect the local magnetism and chemical potential, both of which are crucial for the realization of QAH effect.

**DISSCUSSION**

Our findings shed important lights on better design of magnetic topological insulators. To dilute the magnetic dopant, one efficient way is to add buffer layers between $MnBi_2Te_4$, which has been recently realized in $MnBi_4Te_7$(*39-41*). Another inspiration is the possibility to fine tune magnetic order in related materials. According to the complex layered structures observed in $MnBi_2Te_4$, magnetic interactions are more complicated than expected, which also provides an excellent opportunity to obtain intrinsic ferromagnetic topological insulator.

**METHODS**

**Sample preparation.** To grow $MnBi_2Te_4$ crystals, $Bi_2Te_3$ and MnTe are prepared as precursors by reacting stoichiometric high-purity Bi (99.99%, Adamas) and Te (99.999%, Aladdin), and Mn (99.95%, Alfa Aesar) and Te (99.999%, Aladdin), respectively. The synthesis of $MnBi_2Te_4$ was carried out in a vacuum sealed silica ampoule at 864 K by long-term annealing over one month. This process successfully affords mm-sized shiny crystalline $MnBi_2Te_4$ which were examined with powder X-Ray diffraction using Cu $K_\alpha$ radiation.

**STM characterization.** Our STM experiments were carried out on an ultra-high vacuum (UHV) commercial STM system (Unisoku) which can reach a temperature of 400 mK by using a single-shot $^3$He cryostat. The base pressure for experiment is $1.0 \times 10^{-10}$ Torr. $MnBi_2Te_4$ samples were cleaved *in-situ* at 78 K and then transferred into STM. The bias voltage was applied on samples. The STSs data were obtained by a standard lock-in method that applied an additional small AC voltage with a frequency of 973.0 Hz. The d$I$/d$V$ spectra were collected by disrupting the feedback loop and sweeping the DC bias voltage.

**First-principles calculation.** First-principles calculations were performed by the Vienna *ab initio* Simulation Package(*42*) in the framework of density functional theory. The energy cutoff of plane-wave basis was fixed at 350 eV. The electron-ion interactions were modelled by the projector augmented wave potential, and the Perdew-Burke-Ernzerhof type generalized gradient approximation (GGA)(*43*) was used to approximate the exchange-correlation functional. The GGA + $U$ method was adopted to describe the localized Mn 3$d$-orbitals, and $U$ = 4 eV was selected according to our previous tests(*18*). The structural relaxation, including lattice constants and atomic positions inside the lattice, was performed with a force criterion of 0.01 eV/Å. The DFT-D3 method(*44*) was adopted, given van der Waals interactions between septule layers. The Monkhorst-Pack $k$-point mesh of 9×9×3 was adopted and spin orbit coupling was included in self-consistent calculations. Surface state calculations were performed using WannierTools package(*45*), based on the tight-binding Hamiltonians constructed from maximally localized Wannier functions (MLWF).

**QPI simulation.** The QPI patterns at different energies are simulated by the following scattering possibility formula:

$$P(q, E) = \int dk\, A(k + q, E) A(k, E)$$

where $A(k, E)$ is the spectral function, or the constant energy contours (CECs). The formula above is equivalent to self-auto-correlation method applied on the CECs in $k$ space.

**REFERENCES**

1. M. Z. Hasan, C. L. Kane, Colloquium: Topological insulators. *Rev. Mod. Phys.* **82**, 3045-3067 (2010).

2. X.-L. Qi, S.-C. Zhang, Topological insulators and superconductors. *Rev. Mod. Phys.* **83**, 1057-1110 (2011).

3. F. D. M. Haldane, Model for a quantum Hall effect without Landau levels: condensed-matter realization of the "parity anomaly". *Phys. Rev. Lett.* **61**, 2015-2018 (1988).

4. R. Yu, W. Zhang, H.-J. Zhang, S.-C. Zhang, X. Dai, Z. Fang, Quantized anomalous Hall effect in magnetic topological insulators. *Science* **329**, 61-64 (2010).

5. C.-Z. Chang, J. Zhang, X. Feng, J. Shen, Z. Zhang, M. Guo, K. Li, Y. Ou, P. Wei, L.-L. Wang, Z.-Q. Ji, Y. Feng, S. Ji, X. Chen, J. Jia, X. Dai, Z. Fang, S.-C. Zhang, K. He, Y. Wang, L. Lu, X.-C. Ma, Q.-K. Xue, Experimental observation of the quantum anomalous Hall effect in a magnetic topological insulator. *Science* **340**, 167-170 (2013).

6. J. G. Checkelsky, R. Yoshimi, A. Tsukazaki, K. S. Takahashi, Y. Kozuka, J. Falson, M. Kawasaki, Y. Tokura, Trajectory of the anomalous Hall effect towards the quantized state in a ferromagnetic topological insulator. *Nat. Phys.* **10**, 731 (2014).

7. X. Kou, S.-T. Guo, Y. Fan, L. Pan, M. Lang, Y. Jiang, Q. Shao, T. Nie, K. Murata,

J. Tang, Y. Wang, L. He, T.-K. Lee, W.-L. Lee, K. L. Wang, Scale-invariant quantum anomalous Hall effect in magnetic topological insulators beyond the two-dimensional limit. *Phys. Rev. Lett.* **113**, 137201 (2014).

8. C.-Z. Chang, W. Zhao, D. Y. Kim, H. Zhang, B. A. Assaf, D. Heiman, S.-C. Zhang, C. Liu, M. H. W. Chan, J. S. Moodera, High-precision realization of robust quantum anomalous Hall state in a hard ferromagnetic topological insulator. *Nat. Mater.* **14**, 473-477 (2015).

9. A. J. Bestwick, E. J. Fox, X. Kou, L. Pan, K. L. Wang, D. Goldhaber-Gordon, Precise quantization of the anomalous Hall effect near zero magnetic field. *Phys. Rev. Lett.* **114**, 187201 (2015).

10. S. Grauer, S. Schreyeck, M. Winnerlein, K. Brunner, C. Gould, L. W. Molenkamp, Coincidence of superparamagnetism and perfect quantization in the quantum anomalous Hall state. *Phys. Rev. B* **92**, 201304 (2015).

11. A. M. Essin, J. E. Moore, D. Vanderbilt, Magnetoelectric polarizability and axion electrodynamics in crystalline insulators. *Phys. Rev. Lett.* **102**, 146805 (2009).

12. R. S. K. Mong, A. M. Essin, J. E. Moore, Antiferromagnetic topological insulators. *Phys. Rev. B* **81**, 245209 (2010).

13. F. Katmis, V. Lauter, F. S. Nogueira, B. A. Assaf, M. E. Jamer, P. Wei, B. Satpati, J. W. Freeland, I. Eremin, D. Heiman, P. Jarillo-Herrero, J. S. Moodera, A high-temperature ferromagnetic topological insulating phase by proximity coupling. *Nature* **533**, 513-516 (2016).

14. M. Mogi, M. Kawamura, R. Yoshimi, A. Tsukazaki, Y. Kozuka, N. Shirakawa, K. S. Takahashi, M. Kawasaki, Y. Tokura, A magnetic heterostructure of topological insulators as a candidate for an axion insulator. *Nat. Mater.* **16**, 516-521 (2017).

15. D. Xiao, J. Jiang, J.-H. Shin, W. Wang, F. Wang, Y.-F. Zhao, C. Liu, W. Wu, M. H. W. Chan, N. Samarth, C.-Z. Chang, Realization of the axion insulator state in quantum anomalous Hall sandwich heterostructures. *Phys. Rev. Lett.* **120**, 056801 (2018).

16. W. Li, M. Claassen, C.-Z. Chang, B. Moritz, T. Jia, C. Zhang, S. Rebec, J. J. Lee, M. Hashimoto, D. H. Lu, R. G. Moore, J. S. Moodera, T. P. Devereaux, Z. X. Shen, Origin of the low critical observing temperature of the quantum anomalous Hall effect in V-doped (Bi, Sb)$_2$Te$_3$ film. *Sci. Rep.* **6**, 32732 (2016).

17. Y. Gong, J. Guo, J. Li, K. Zhu, M. Liao, X. Liu, Q. Zhang, L. Gu, L. Tang, X. Feng, D. Zhang, W. Li, C. Song, L. Wang, P. Yu, X. Chen, Y. Wang, H. Yao, W. Duan, Y. Xu, S.-C. Zhang, X. Ma, Q.-K. Xue, K. He, Experimental realization of an intrinsic magnetic topological insulator. *Chin. Phys. Lett.* **36**, 076801 (2019).

18. J. Li, Y. Li, S. Du, Z. Wang, B.-L. Gu, S.-C. Zhang, K. He, W. Duan, Y. Xu, Intrinsic magnetic topological insulators in van der Waals layered $MnBi_2Te_4$-family materials. *Sci. Adv.* **5**, eaaw5685 (2019).

19. D. Zhang, M. Shi, T. Zhu, D. Xing, H. Zhang, J. Wang, Topological axion states in the magnetic insulator $MnBi_2Te_4$ with the quantized magnetoelectric effect. *Phys. Rev. Lett.* **122**, 206401 (2019).

20. Y. Deng, Y. Yu, M. Shi, J. Wang, X. Chen, Y. Zhang, Magnetic-field-induced quantized anomalous Hall effect inintrinsic magnetic topological insulator $MnBi_2Te_4$. *arXiv: 1904.11468*, (2019).

21. C. Liu, Y. Wang, H. Li, Y. Wu, Y. Li, J. Li, K. He, Y. Xu, J. Zhang, Y. Wang, Quantum phase transition from axion insulator to Chern insulator in $MnBi_2Te_4$. *arXiv: 1905.00715*, (2019).

22. J. Ge, Y. Liu, J. Li, H. Li, T. Luo, Y. Wu, Y. Xu, J. Wang, High-Chern-number and high-temperature quantum Hall effect without Landau levels. *arXiv: 1907.09947*, (2019).

23. S. H. Lee, Y. Zhu, Y. Wang, L. Miao, T. Pillsbury, H. Yi, S. Kempinger, J. Hu, C. A. Heikes, P. Quarterman, W. Ratcliff, J. A. Borchers, H. Zhang, X. Ke, D. Graf, N. Alem, C.-Z. Chang, N. Samarth, Z. Mao, Spin scattering and noncollinear spin structure-induced intrinsic anomalous Hall effect in antiferromagnetic topological insulator $MnBi_2Te_4$. *Phys. Rev. Research* **1**, 012011 (2019).

24. B. Chen, F. Fei, D. Zhang, B. Zhang, W. Liu, S. Zhang, P. Wang, B. Wei, Y. Zhang, Z. Zuo, J. Guo, Q. Liu, Z. Wang, X. Wu, J. Zong, X. Xie, W. Chen, Z. Sun, S. Wang, Y. Zhang, M. Zhang, X. Wang, F. Song, H. Zhang, D. Shen, B. Wang, Intrinsic magnetic topological insulator phases in the Sb doped $MnBi_2Te_4$ bulks and thin flakes. *Nature Commun.* **10**, 4469 (2019).

25. R. C. Vidal, H. Bentmann, T. R. F. Peixoto, A. Zeugner, S. Moser, C. H. Min, S. Schatz, K. Kißner, M. Ünzelmann, C. I. Fornari, H. B. Vasili, M. Valvidares, K. Sakamoto, D. Mondal, J. Fujii, I. Vobornik, S. Jung, C. Cacho, T. K. Kim, R. J.

Koch, C. Jozwiak, A. Bostwick, J. D. Denlinger, E. Rotenberg, J. Buck, M. Hoesch, F. Diekmann, S. Rohlf, M. Kalläne, K. Rossnagel, M. M. Otrokov, E. V. Chulkov, M. Ruck, A. Isaeva, F. Reinert, Surface states and Rashba-type spin polarization in antiferromagnetic MnBi$_2$Te$_4$. *Phys. Rev. B* **100**, 121104 (2019).

26. Y.-J. Hao, P. Liu, Y. Feng, X.-M. Ma, E. F. Schwier, M. Arita, S. Kumar, C. Hu, R. e. Lu, M. Zeng, Y. Wang, Z. Hao, H.-Y. Sun, K. Zhang, J. Mei, N. Ni, L. Wu, K. Shimada, C. Chen, Q. Liu, C. Liu, Gapless Surface Dirac Cone in Antiferromagnetic Topological Insulator MnBi$_2$Te$_4$. *Phy. Rev. X* **9**, 041038 (2019).

27. Y. J. Chen, L. X. Xu, J. H. Li, Y. W. Li, H. Y. Wang, C. F. Zhang, H. Li, Y. Wu, A. J. Liang, C. Chen, S. W. Jung, C. Cacho, Y. H. Mao, S. Liu, M. X. Wang, Y. F. Guo, Y. Xu, Z. K. Liu, L. X. Yang, Y. L. Chen, Topological Electronic Structure and Its Temperature Evolution in Antiferromagnetic Topological Insulator MnBi$_2$Te$_4$. *Phy. Rev. X* **9**, 041040 (2019).

28. P. Swatek, Y. Wu, L.-l. Wang, K. Lee, B. Schrunk, J. Yan, A. Kaminski, Gapless Dirac surface states in the antiferromagnetic topological insulator MnBi$_2$Te$_4$. *arXiv: 1907.09596*, (2019).

29. I. Lee, C. K. Kim, J. Lee, S. J. L. Billinge, R. Zhong, J. A. Schneeloch, T. Liu, T. Valla, J. M. Tranquada, G. Gu, J. C. S. Davis, Imaging Dirac-mass disorder from magnetic dopant atoms in the ferromagnetic topological insulator Cr$_x$(Bi$_{0.1}$Sb$_{0.9}$)$_{2-x}$Te$_3$. *Proc. Natl. Acad. Sci. U.S.A.* **112**, 1316-1321 (2015).

30. T. Zhang, P. Cheng, X. Chen, J.-F. Jia, X. Ma, K. He, L. Wang, H. Zhang, X. Dai, Z. Fang, X. Xie, Q.-K. Xue, Experimental demonstration of topological surface states protected by time-reversal symmetry. *Phys. Rev. Lett.* **103**, 266803 (2009).

31. B. Huang, G. Clark, E. Navarro-Moratalla, D. R. Klein, R. Cheng, K. L. Seyler, D. Zhong, E. Schmidgall, M. A. McGuire, D. H. Cobden, W. Yao, D. Xiao, P. Jarillo-Herrero, X. Xu, Layer-dependent ferromagnetism in a van der Waals crystal down to the monolayer limit. *Nature* **546**, 270-273 (2017).

32. J. Cui, M. Shi, H. Wang, F. Yu, T. Wu, X. Luo, J. Ying, X. Chen, Transport properties of thin flakes of the antiferromagnetic topological insulator MnBi$_2$Te$_4$. *Phys. Rev. B* **99**, 155125 (2019).

33. A. Zeugner, F. Nietschke, A. U. B. Wolter, S. Gaß, R. C. Vidal, T. R. F. Peixoto, D. Pohl, C. Damm, A. Lubk, R. Hentrich, S. K. Moser, C. Fornari, C. H. Min, S.

Schatz, K. Kißner, M. Ünzelmann, M. Kaiser, F. Scaravaggi, B. Rellinghaus, K. Nielsch, C. Hess, B. Büchner, F. Reinert, H. Bentmann, O. Oeckler, T. Doert, M. Ruck, A. Isaeva, Chemical Aspects of the Candidate Antiferromagnetic Topological Insulator $MnBi_2Te_4$. *Chem. Mater.* **31**, 2795-2806 (2019).

34. J.-Q. Yan, Q. Zhang, T. Heitmann, Z. Huang, K. Chen, J.-G. Cheng, W. Wu, D. Vaknin, B. C. Sales, R. J. McQueeney, Crystal growth and magnetic structure of $MnBi_2Te_4$. *Phys. Rev. M* **3**, 064202 (2019).

35. M. M. Otrokov, I. I. Klimovskikh, H. Bentmann, D. Estyunin, A. Zeugner, Z. S. Aliev, S. Gaß, A. U. B. Wolter, A. V. Koroleva, A. M. Shikin, M. Blanco-Rey, M. Hoffmann, I. P. Rusinov, A. Y. Vyazovskaya, S. V. Eremeev, Y. M. Koroteev, V. M. Kuznetsov, F. Freyse, J. Sánchez-Barriga, I. R. Amiraslanov, M. B. Babanly, N. T. Mamedov, N. A. Abdullayev, V. N. Zverev, A. Alfonsov, V. Kataev, B. Büchner, E. F. Schwier, S. Kumar, A. Kimura, L. Petaccia, G. Di Santo, R. C. Vidal, S. Schatz, K. Kißner, M. Ünzelmann, C. H. Min, S. Moser, T. R. F. Peixoto, F. Reinert, A. Ernst, P. M. Echenique, A. Isaeva, E. V. Chulkov, Prediction and observation of an antiferromagnetic topological insulator. *Nature* **576**, 416-422 (2019).

36. P. Cheng, C. Song, T. Zhang, Y. Zhang, Y. Wang, J. F. Jia, J. Wang, Y. Wang, B. F. Zhu, X. Chen, X. Ma, K. He, L. Wang, X. Dai, Z. Fang, X. Xie, X. L. Qi, C. X. Liu, S. C. Zhang, Q. K. Xue, Landau quantization of topological surface states in $Bi_2Se_3$. *Phys. Rev. Lett.* **105**, 076801 (2010).

37. T. Hanaguri, K. Igarashi, M. Kawamura, H. Takagi, T. Sasagawa, Momentum-resolved Landau-level spectroscopy of Dirac surface state in $Bi_2Se_3$. *Phys. Rev. B* **82**, 081305(R) (2010).

38. J. Li, C. Wang, Z. Zhang, B.-L. Gu, W. Duan, Y. Xu, Magnetically controllable topological quantum phase transitions in the antiferromagnetic topological insulator $MnBi_2Te_4$. *Phys. Rev. B* **100**, 121103 (2019).

39. C. Hu, X. Zhou, P. Liu, J. Liu, P. Hao, E. Emmanouilidou, H. Sun, Y. Liu, H. Brawer, A. P. Ramirez, H. Cao, Q. Liu, D. Dessau, N. Ni, A van der Waals antiferromagnetic topological insulator with weak interlayer magnetic coupling. *arXiv: 1905.02154*, (2019).

40. J. Wu, F. Liu, M. Sasase, K. Ienaga, Y. Obata, R. Yukawa, K. Horiba, H. Kumigashira, S. Okuma, T. Inoshita, H. Hosono, Natural van der Waals heterostructural single crystals with both magnetic and topological properties. *Sci. Adv.* **5**, eaax9989 (2019).

41. R. C. Vidal, A. Zeugner, J. I. Facio, R. Ray, M. H. Haghighi, A. U. Wolter, L. T. C. Bohorquez, F. Caglieris, S. Moser, T. Figgemeier, T. R. F. Peixoto, H. B. Vasili, M. Valvidares, S. Jung, C. Cacho, A. Alfonsov, K. Mehlawat, V. Kataev, C. Hess, M. Richter, B. Büchner, J. van den Brink, M. Ruck, F. Reinert, H. Bentmann, A. Isaeva, Topological electronic structure and intrinsic magnetization in $MnBi_4Te_7$: a $Bi_2Te_3$-derivative with a periodic Mn sublattice. *arXiv: 1906.08394,* (2019).

42. G. Kresse, J. Furthmüller, Efficient iterative schemes for ab initio total-energy calculations using a plane-wave basis set. *Phys. Rev. B* **54**, 11169 (1996).

43. J. P. Perdew, K. Burke, M. Ernzerhof, Generalized gradient approximation made simple. *Phys. Rev. Lett.* **77**, 3865 (1996).

44. S. Grimme, J. Antony, S. Ehrlich, H. Krieg, A consistent and accurate ab initio parametrization of density functional dispersion correction (DFT-D) for the 94 elements H-Pu. *J. Chem. Phys.* **132**, 154104 (2010).

45. Q. Wu, S. Zhang, H.-F. Song, M. Troyer, A. A. Soluyanov, WannierTools: An open-source software package for novel topological materials. *Comp. Phys. Commun.* **224**, 405-416 (2018).

**Acknowledgements**

**Funding:** The experimental work was supported by the National Science Foundation (No. 11674191, No. 11674165, No. U1832218), Ministry of Science and Technology of China (No. 2016YFA0301002) and the Beijing Advanced Innovation Center for Future Chip (ICFC), National key R&D Program of China (Grant 2016YFB0901600), Science and Technology Commission of Shanghai (Grant 16JC1401700), CAS Center for Excellence in Superconducting Electronics. W. Li was also supported by Beijing Young Talents Plan and the National Thousand-Young-Talents Program. **Author Contributions:** W.L. and Q-K.X. conceived and designed the experiment. Y.Y. and X.W. performed the STM experiments. H.L. and Y.W. grew the samples. J.L., Y.X. and W.D. carried out the DFT calculations. Y.J., Z.H., Y.W. K.H. analyzed the data. W.L. and Y.Y. wrote the manuscript with inputs from all other authors. We thank Lexian Yang, Haijun Zhang and Yulin Chen for helpful discussions. **Competing interests:** The authors declare that they have no competing interest. **Data and materials availability:** All data needed to evaluate the conclusions in the paper are present in the paper. Additional data related to this paper may be requested from the authors.

**FIGURES AND CAPTIONS**

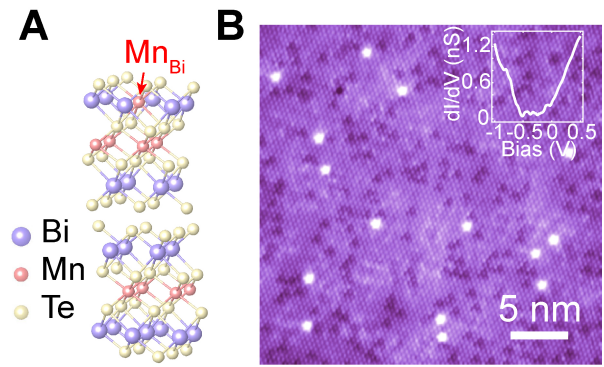

**Fig. 1. Lattice and electronic structures of MnBi$_2$Te$_4$.** (A) Lattice structure of MnBi$_2$Te$_4$. A Mn$_{Bi}$ antisite defect is demonstrated in the second atomic layer. (B) Atomically resolved topographic image (30 nm × 30 nm, bias voltage $V_s$ = -1.2 V, tunneling current $I_t$ = 100 pA) of MnBi$_2$Te$_4$. It shows hexagonal lattice with lattice constant of $a$ = 4.33 Å. Two types of defects, bright dots and dark triangles, are revealed on the Te-terminated (111) surface and they probably correspond to adatoms and Mn$_{Bi}$, respectively. Inset: d$I$/d$V$ spectrum obtained by averaging 2601 spectra taken in a 25 nm × 25 nm area (set point: $V_s$ = 500 mV, $I_t$ = 400 pA).

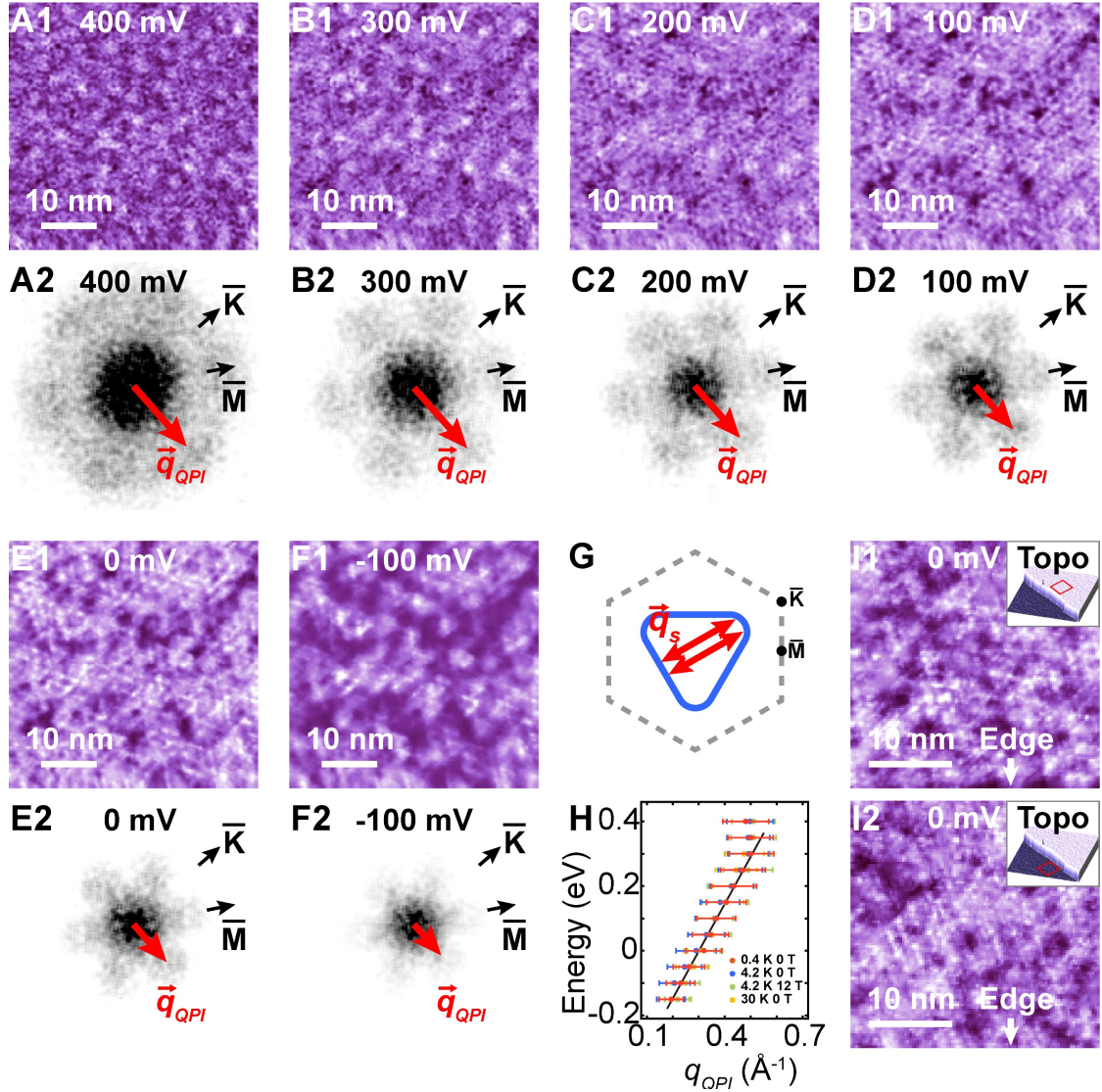

**Fig. 2. Quasiparticle interference (QPI) results on MnBi$_2$Te$_4$.** (**A1-F1**) The real space d$I$/d$V$ maps taken at various bias voltages (45 nm × 45 nm, set point: $V_s$ = 800 mV, $I_t$ = 800 pA). (**A2-F2**) Fast Fourier transformation of the real space d$I$/d$V$ maps. The red arrows denote the extracted scattering wave vectors $\vec{q}_{QPI}$. (**G**) Schematic of three-fold symmetric surface state (blue line) in MnBi$_2$Te$_4$ with $z$-direction A-type antiferromagnetic order. The double-headed red arrows denote the dominant quasiparticle scattering wave vector related with surface states (along $\overline{\Gamma}$-$\overline{K}$ direction). (**H**) Energy dispersions of the scattering wave vectors $\vec{q}_{QPI}$ at elevated temperature and 12 T magnetic field. The error bar denotes the radius of the spot patterns in the corresponding FFT images. No significant difference is observed between the scattering wave vectors obtained from different experimental conditions. The black line is the linear fit to the data taken at 0.4 K, 0 T. (**I1, I2**) d$I$/d$V$ maps taken on upper and lower terraces, respectively (30 nm × 30 nm, set point: $V_s$ = 800 mV, $I_t$ = 800 pA). Standing

wave is absent near the terrace edges. Insets: The corresponding topographic images of the terraces. The red squares denote the mapping regions.

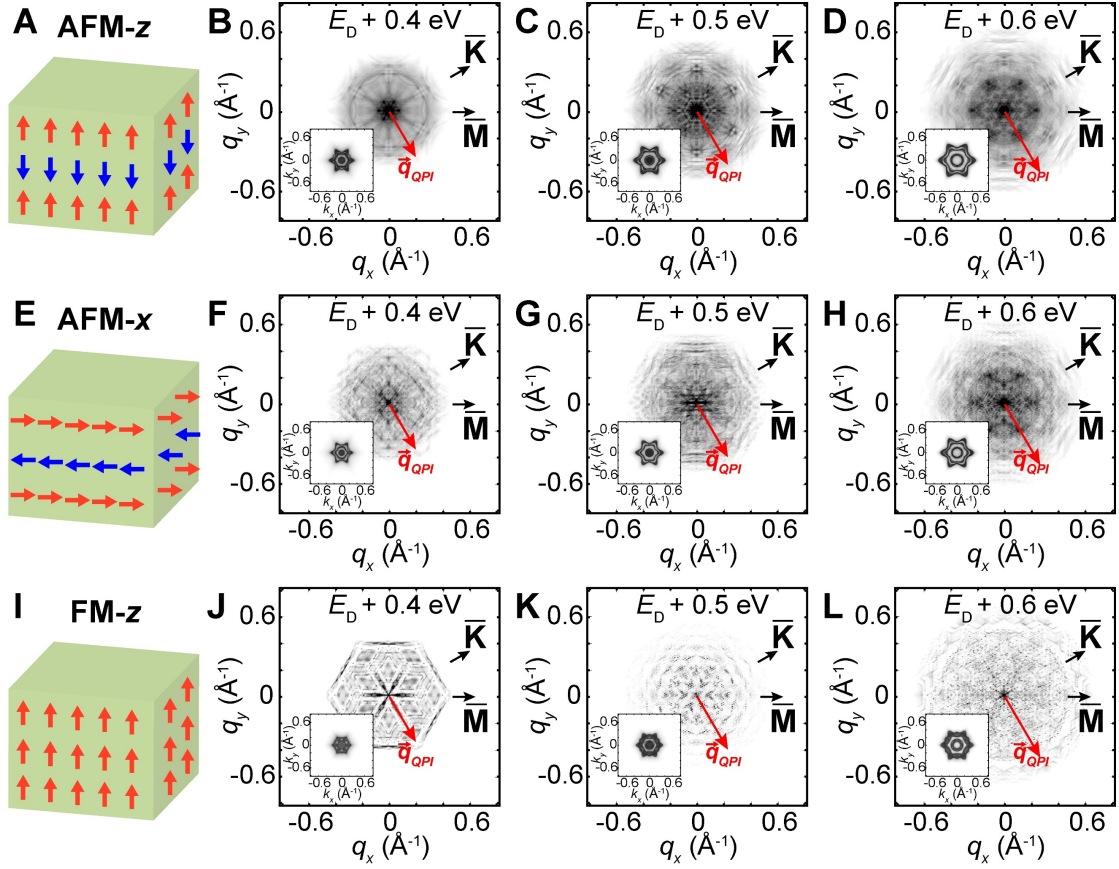

**Fig. 3. Simulated QPI patterns by CECs of MnBi$_2$Te$_4$ with different magnetisms.** (**A**) Schematic of out-of-plane A-type AFM order (AFM-$z$) in MnBi$_2$Te$_4$. (**B-D**) The simulated interference patterns with out-of-plane A-type AFM order at $E_D$ + 0.4 eV, 0.5 eV and 0.6 eV, respectively. Here, $E_D$ denotes the Dirac point energy and we take $E_D$ = -0.27 eV. The red arrows denote the scattering vectors $\vec{q}_{QPI}$ obtained from the fitting result in Fig. 2H. Insets: The total CECs obtained from DFT calculations at different energies. (**E**) Schematic of in-plane A-type AFM order (AFM-$x$) in MnBi$_2$Te$_4$. (**F-H**) The simulated interference patterns with in-plane A-type AFM order. (**I**) Schematic of out-of-plane ferromagnetic order (FM-$z$) in MnBi$_2$Te$_4$. (**J-L**) The simulated interference patterns with FM order.

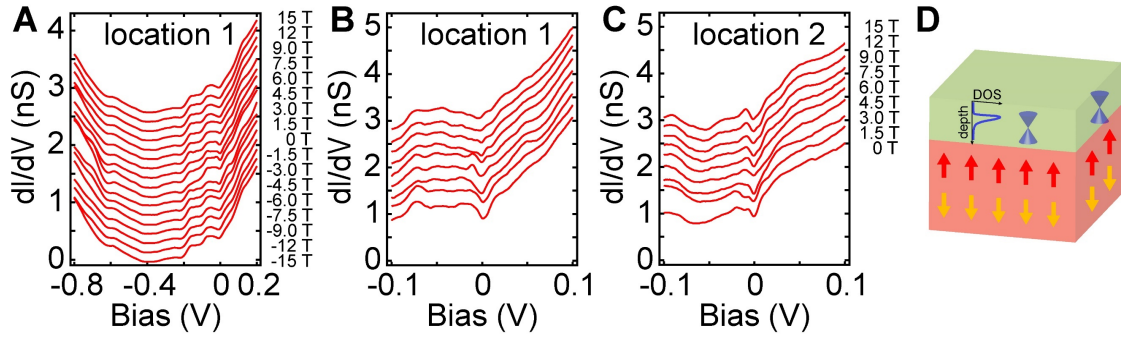

**Fig. 4. Magnetic response in the d$I$/d$V$ spectra.** (**A**) Large-range d$I$/d$V$ spectra taken at a specific defect-free location with applying various magnetic fields (set point: $V_s$ = 200 mV, $I_t$ = 200 pA). Each spectrum is shifted for clarity. (**B** and **C**) Small-range d$I$/d$V$ spectra taken at defect-free locations with applying various magnetic fields (set point: $V_s$ = 100 mV, $I_t$ = 200 pA). Some fluctuating features are presented near $E_F$ but with no systematic evolution to magnetic fields. (**D**) A possible layered structure that account for the experimental observations in STM, ARPES and transport measurements. The top layer with green color hosts an embedded Dirac type surface state; And the bottom red bulk layer is the predicted out-of-plane antiferromagnetic bulk.

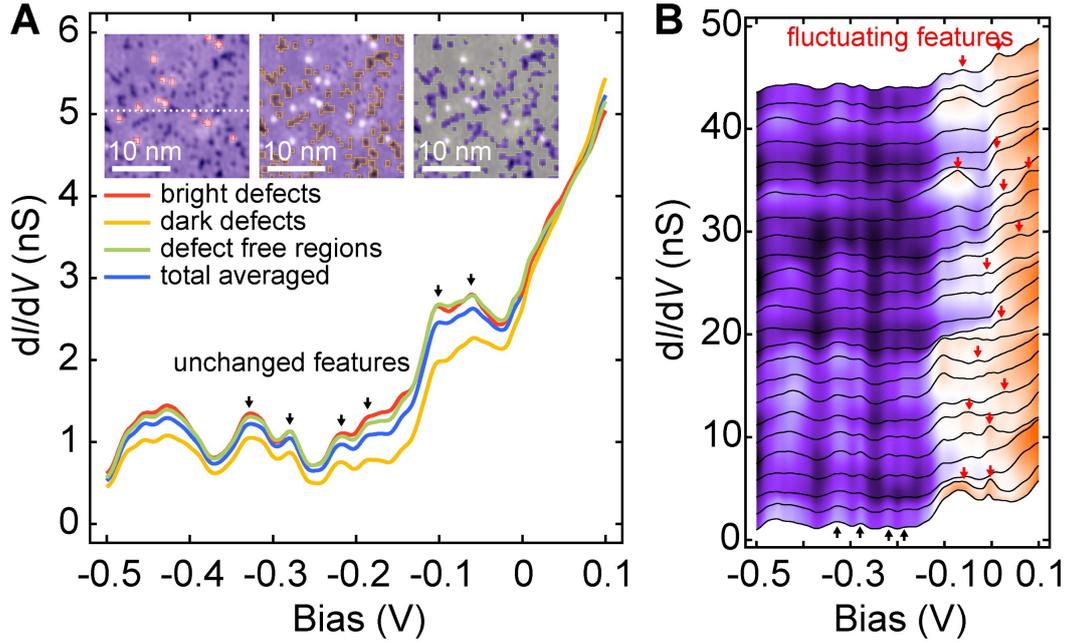

**Fig. 5. Locally fluctuating features in d$I$/d$V$ spectra.** (**A**) Averaged d$I$/d$V$ spectra taken on the bright defects, dark triangles and defect-free regions (set point: $V_s$ = 100 mV, $I_t$ = 400 pA). Insets: The corresponding positions in which the d$I$/d$V$ spectra are taken to calculate of the averaged spectra. The red, orange and green shadowed regions correspond to the bright defects, dark triangles and defect-free regions, respectively. (**B**) A series of d$I$/d$V$ spectra (set point, $V_s$ = 100 mV, $I_t$ = 400 pA) taken along the white dotted line in the inset of (**A**). The DOS show fluctuating features near $E_F$.